\documentstyle[12pt]{article}
\begin{document}
\newpage
\begin{center}
\Large {EFFECTS OF DICKE SUPERRADIANCE IN THE}\\
\end{center}
\vskip 0.4cm
\begin{center} 
\Large{CONTEXT OF THE ONE-ATOM MASER}\\
\end{center}
\vskip 0.8cm
\begin{center}
\large{
N. Nayak$^a$, A. S. Majumdar$^b$, and V. Bartzis$^c$} \\
\end{center}
\vskip 0.2in

\begin{center}
$^{a \& b}$S. N. Bose National Centre for Basic Sciences,\\ 
\end{center}
\begin{center}
Block-JD, Sector-3, Salt Lake City, Calcutta-700091, India \\
\end{center}
\vskip 0.2in

\begin{center}
$^c$General Department of Physics, Chemistry and Material Technology,\\
\end{center}
\begin{center}
Technological Educational Institutions of Athens, Egaleo 12210, Greece\\
\end{center}

%\newpage
%\baselineskip=24pt
%\date
%\maketitle
\vskip 0.5in
\begin{center}
\bf {Abstract}\\
\end{center}
\indent
We consider a micromaser model to study the influence of Dicke
superradiance in the context of the one-atom maser. The model
involves a microwave 
cavity into which two-level Rydberg atoms are pumped in pairs. We
consider a  random pump mechanism which allows the presence of at most one pair 
of atoms in the cavity at any time.
We analyze the
differences between the present system, called the Dicke micromaser,  
and an equivalently pumped
conventional one-atom micromaser. These differences are attributed to
the Dicke cooperativity in the two-atom system.
We also show that the two-atom Dicke
micromaser is equivalent to a one-atom cascade two-photon micromaser. 
With the introduction of a one-photon detuning, the present theory further
describes a true two-photon micromaser. We discuss in detail the role of
one-photon detuning in the mechanism of a one-atom two-photon micromaser.
This leads us to point out that the two-atom cavity dynamics can be verified
by a proper scaling of the results from an equivalent 
one-atom two-photon micromaser. 
\baselineskip=24pt
\vskip 0.5in
PACS Nos  : 42.50.Dv, 42.52.+x, 32.80.-t     \\ \\
\newpage  
\noindent
1. Introduction

\vskip 0.2in
 
The first experimental observation of maser action with just one atom in the
so-called micromaser~[1] renewed interest in the subject extensively as it
opened possibilities of observing cavity-QED effects which are purely of
quantum mechanical origin. Indeed, one of the results of the much discussed
Jaynes-Cummings interactions~[2], the quantum revival 
of the oscillations in the
atomic population, has been observed in the micromaser device~[3]. The
micromaser device~[1] consists of a high-Q superconducting cavity cooled down
to sub-Kelvin temperatures into which Rydberg atoms in the upper of the
two masing levels are pumped randomly but one at a time. A clever velocity
selector decides a fixed flight time $\tau$ through the cavity for  every
 atom. The repetition time $\bar{t}_{R}$, $\tau$ and $t_{cav}$, the duration
in which the cavity is empty of atom, satisfy $\bar{t}_{R}=\tau + t_{cav}$
with $\tau \ll t_{cav}$. This is the basis of the one-atom maser
theories~[4-6] with the further assumption of strictly single-atom events.

However, the above arrangements in the experiments~[1,3] cannot eliminate
overlap of flight times through the cavity of successive atoms completely,
which of course, is less than $1\%$. This has been one of the obstacles 
in realizing
certain predictions of one-atom maser theories~[4,5] such as number states
of the cavity radiation field. Another limiting factor is the dissipative 
mechanism acompanying the coherent atom-field interaction~[6]. The results
in [6] show that the effects of the dissipative mechanism can be minimized
by increasing the $Q$ factor of the cavity and by decreasing its temperature.
But the pump mechanism has not been able to eliminate traces of two atom
events in one-atom maser dynamics. Hence it has become necessary to 
explain the effect of such two-atom events on the photon statistics of the
cavity field. Various models incorporating multiple atomic events have
been proposed~[7--13]. The basic difference between such models and 
one-atom maser theory~[4-6] is the Dicke atomic cooperative effect~[14,15]
acompanying the multiple atomic events. Thus, in order to understand the
limitations of maser theories based on strictly single atom events~[4-6], it is
necessary to quantify the effect of atomic cooperation. For this purpose
we propose the following model.

We consider a fixed number of atoms entering the cavity with a constant
flight time through it, but with random 
time gaps between successive such events.
The cooperative nature~[12] of the interaction of the atoms with the
cavity field shall be fully reflected in the photon statistics. In
this paper we consider atoms being pumped in pairs into the cavity.
We limit the number of atoms to two just
for the sake of simplicity which can be easily generalized to
 a larger number. Both the atoms are taken to be in their upper states
 when they enter the cavity. The arrival of the pair at the cavity
 is assumed to be Poissonian. 
 We thus have ${\bar t}_{R}=\tau +{\bar t}_{cav}$ with $t_{R}=1/\bar{R}$. 
We assume
$\tau$ to be a fixed duration for every pair of atoms, and hence,
$t_{cav}$ is random.  
$\bar{R}$
 is now a Poisson average of R pairs of atoms pumped into the cavity
 in one second. Due to cooperative interaction of the pair of atoms with
 the cavity field, we will find its dynamics with $\bar{R}$ pairs of
 atoms pumped into the cavity per second in the present case being different
 from the one-atom maser pumped with $2\bar{R}$ atoms into the cavity 
per second. This is due to
the superradiance 
effects~[14,15] which will be manifested in the steady-state photon statistics. 
For these reasons, we call the present system a ``Dicke micromaser''. 

It is rather a difficult task to achieve  such a model experimentally.
However, we will show in the following that the present model is
closely related to a one-atom two-photon micromaser~[16-18]. 
This is a key result of this paper. Two-photon micromaser action with one
atom has been experimentally demonstrated~[16]. Thus
the model we are considering to show the atomic cooperative effects
is not beyond experimental verification.

\indent
The organization of the paper is as follows. In the section 2, we present the
 microscopic Hamiltonian and the Dicke states involved in the micromaser 
action. The equations of motion are derived in section 3. We derive and
discuss the photon statistics  in section 4.  
The present system is compared with the 
two-photon micromasers in section 5. We
 conclude the paper in section 6.\\

\vskip 0.5in

\noindent
2. The Model

\vskip 0.2in

A pair of Rydberg atoms in the upper of their two levels $a$ and $b$
enter the cavity at $t = 0$. We assume that the atomic transition frequency
is in resonance with the cavity mode frequency $\omega$. 
Both the atoms take the time $\tau$
to travel across the cavity. During this time, the evolution of the 
system comprising of the two atoms and the single eigen mode of the 
microwave cavity is governed by the Hamiltonian
$$H = H_0 + H_I \eqno(1a)$$
where
$$H_0 = \omega (S^Z_1 + S^Z_2 + a^{\dagger}a)\eqno(1b)$$
as we consider resonance condition between atomic transition frequencies
and the cavity frequency. $H_I$ is the Jaynes-Cummings Hamiltonian~[2].
$$H_I = g \sum_{i=1}^2 (S^{+}_ia + S^{-}_ia^{\dagger})\eqno(1c)$$
We assume that the atom-field coupling constant $g$ is the same for
each atom. The atomic operators obey the relations
$[S^{+}_i, S^{-}_j] = 2S^Z_i\delta_{ij}$ and $a(a^{\dagger})$ is the
photon annihilation (creation) operator obeying the commutation
relation $[a, a^{\dagger}] = 1$. In the frame rotating at $H_0$, we
have $H= H_I$.

It is convenient to represent the two-atom system by a Dicke system
as both the atoms interact with the same cavity field simultaneously.
In doing so, we however, neglect the dipole-dipole interaction as the 
microwave cavity dimensions are much large compared to particle 
wavelengths. Defining the collective operators $S^{\pm} =
\sum_{i=1}^2 S^{\pm}_i$, Eq.(1c) takes the form
$$H_I = g(S^{+}a + S^{-}a^{\dagger})\eqno(2)$$
which is clearly invariant by atomic permutation. Both the atoms are in
their respective upper states at $t=0$ and hence, the state representing
the atomic system $|a,a\rangle$ is also invariant under atomic
permutation. So it remains in a symmetrical state at all times. These
are known as Dicke states with maximum cooperative number $J=1$ (we have
a two-atom system) and these states are isomorphous to angular momentum
states with principal angular momentum $J=1$. These states, represented
by $|J,M\rangle$, are generated by
$$S^{-}|J,M\rangle = \sqrt{(J+M)(J-M+1)}|J,M-1\rangle$$
and 
$$\eqno(3)$$
$$S^{+}|J,M\rangle = \sqrt{(J+M+1)(J-M)}|J,M+1\rangle$$
The level $|J,M\rangle$ is nondegenerate and has atomic energy $M\omega$.
As $-J \le M \le J$, this has $2J+1$ states, and in the present case,
the number of states is three. Since $J$ is a constant here, we
represent $|J,M\rangle \equiv |M\rangle$.

Thus, the pair of two-level atoms can be represented by a three-level
system of the Dicke states $|M=1\rangle$, $|M=0\rangle$, and $|M=-1\rangle$,
having energy $\omega$, $0$, and $-\omega$ respectively. This allows us
to write the Hamiltonian in Eq.(2) in the matrix form
$$H_I = \pmatrix{0 & g\sqrt{2(n-1)} & 0 \cr g\sqrt{2(n-1)} & 0 & g\sqrt{2n}
\cr 0 & g\sqrt{2n} & 0}\eqno(4)$$
where $n$ represents the photon number of the cavity field. Eq.(4) resembles
the case for a two-photon process in a three-level system~[19]. Following
the method in [19], we can obtain the eigenvalues and eigenstates of
the interaction of the Dicke states with the cavity radiation field. The
eigenvalues
$$\lambda_0 = 0,$$
$$\lambda_{+} = g\sqrt{2(2n-1)},\eqno(5)$$
$$\lambda_{-} = -g\sqrt{2(2n-1)},$$
and its states are, respectively,
$$|0,n\rangle = x_1^{(n)}|1,n-2\rangle + y_1^{(n)}|0,n-1\rangle +
z_1^{(n)}|-1,n\rangle,$$
$$|+,n\rangle = x_2^{(n)}|1,n-2\rangle + y_2^{(n)}|0,n-1\rangle +
z_2^{(n)}|-1,n\rangle,\eqno(6)$$
$$|-,n\rangle = x_3^{(n)}|1,n-2\rangle + y_3^{(n)}|0,n-1\rangle +
z_3^{(n)}|-1,n\rangle,$$
where $|1,n\rangle$, $|0,n\rangle$ and $|-1,n\rangle$ are the composite
Dicke and field states, and
$$x_1^{(n)} = \sqrt{{n \over 2n-1}} , \>\> y_1^{(n)} = 0 , \>\> z_1^{(n)} = 
-\sqrt{{n-1 \over 2n-1}},\eqno(7a)$$
$$x_2^{(n)} = -\sqrt{{n-1 \over 4n-2}} , \>\> 
y_2^{(n)} = -{1 \over \sqrt{2}}
, \>\> z_2^{(n)} =
-\sqrt{{n \over 4n-2}},\eqno(7b)$$
$$x_3^{(n)} = -\sqrt{{n-1 \over 
4n-2}} , 
\>\> y_3^{(n)} =
{l \over \sqrt{2}}  , \>\> z_3^{(n)} =
-\sqrt{{n \over 4n-2}}.\eqno(7c)$$
The states $|+,n\rangle$, $|-,n\rangle$ and $|0,n\rangle$ can be called
as dressed states of the interaction of the Dicke states with the cavity field.

\vskip 0.5in

\noindent
3. Derivation of the equations of motion

\vskip 0.2in

Using the dressed states in Eq.(6) as the basic states, the equation of motion
$$\dot{\rho} = -i[H, \rho]\eqno(8)$$
can be derived easily~[19]. $\rho$ represents the composite atom-field
system having the initial conditions
$$\rho^{(n)}_{i,j}(t = 0) \equiv \langle i, n|\rho |j, n\rangle =
x^{(n)}_ix^{(n)}_jP_{n-2}\eqno(9)$$
where $P_n$ is the photon distribution function of the cavity radiation
field. In writing Eq.(6), we have neglected the influences of the atomic as
well as the cavity reservoirs on the dynamics. These effects can play a
crucial role in the one-atom micromaser~[6]. However, these influences
have been shown to be negligible~[6] for a cavity having very high $Q$ and
at very low temperatures and a low atomic pump rate. Eq.(8) represents  
such a  system.

The time-dependent solutions for the density matrix elements can be written as
$$\rho^{(n)}_{i,j}(t) = P_{n-2} x_i^{(n)} x_j^{(n)} 
e^{-i(\lambda_i^{(n)} - \lambda_j^{(n)})t},\eqno(10)$$
$$i, j \> \> \equiv 0, +, -.$$
By inverting the matrix which expresses the dressed states as a linear
superposition of the Dicke states, we can obtain the Dicke states
$|i,n\rangle$, $i=1,0,-1$, in terms of the dressed states. Using these
relations and Eq.(10),  we obtain the density matrix elements in the Dicke
state basis. We follow these steps to get the $P_n$ at $t = \tau$, which is
given by the trace
$$P_n(\tau) = \sum_{M=-1}^{+1}\langle M,n|\rho |M,n\rangle.\eqno(11)$$
We thus have
$$P_n(\tau) = \Theta_1^{(n)}(\tau)P_n + \Theta_2^{(n)}(\tau)P_{n-1} + 
\Theta_3^{(n)}(\tau)P_{n-2}\eqno(12)$$
where
$$\Theta_1^{(n)}(\tau) = {(n+2) \over (4n+3)}(x_1^{(n+2)})^2 + {(n+1) 
\over (4n+6)}\biggl(
(x_2^{(n+2)})^2 + (x_3^{(n+2)})^2\biggr)$$ 
$$ - 
{2\sqrt{(n+2)(n+1)} \over \sqrt{(4n+3)(4n+6)}}\biggl(x_1^{(n+2)}x_2^{(n+2)} Re.
(e^{i\lambda_{+}^{(n+2)}\tau}) 
+  x_1^{(n+2)}x_3^{(n+2)} Re.
(e^{i\lambda_{-}^{(n+2)}\tau})\biggr)$$ 
$$+ {2(n+1) \over (4n+6)}x_2^{(n+2)}x_3^{(n+2)} Re.
(e^{-i(\lambda_{+}^{(n+2)}-\lambda_{-}^{(n+2)})\tau}),\eqno(13a)$$

$$\Theta_2^{(n)}(\tau) = {1\over 2}\biggl((x_2^{(n+1)})^2 + 
(x_3^{(n+1)})^2\biggr) - x_2^{(n+1)}x_3^{(n+1)} Re.
(e^{-i(\lambda_{+}^{(n+1)} - \lambda_{-}^{(n+1)})\tau}),\eqno(13b)$$

$$\Theta_3^{(n)}(\tau) = {n-1 \over 2n-1}(x_1^{(n)})^2 +
{n \over 4n-2}\biggl((x_2^{(n)})^2 + (x_3^{(n)})^2\biggr)$$ 
$$+ {\sqrt{2n(n-1)} \over (2n-1)}\biggl(x_1^{(n)}x_2^{(n)} Re.
(e^{i\lambda_{+}^{(n)}\tau}) + x_1^{(n)}x_3^{(n)}
Re.(e^{i\lambda_{-}^{(n)}\tau})\biggr)$$ 
$$+ {2n \over 4n-2}x_2^{(n)}x_3^{(n)} Re.(e^{-i(\lambda_{+}^{(n)} -
\lambda_{-}^{(n)})\tau}).\eqno(13c)$$
The change in $P_n$ at $t=\tau$ is then $\delta P_n = P_n(\tau) - P_n$
where $P_n$ is the photon distribution function at the time of a pair of
atoms entering the cavity. For a time $\triangle t$ such that
$\tau \ll \triangle t \ll t_p$, where $t_p$ is the cavity photon lifetime,
we have
$$\triangle P_n = \delta P_n R \triangle t\eqno(14)$$
where $R$ is the number of pairs of atoms passing through the cavity in
one second

The coarse-grained time derivative due to gain from the atomic interaction
is given by
$${dP_n \over dt}|_{gain} = R\biggl((\Theta_1^{(n)}-1)P_n + \Theta_2^{(n)}
P_{n-1} + \Theta_3^{(n)}P_{n-2}\biggr).\eqno(15)$$
Eq.(15) represents the dynamics during time $\tau$. During the time lapse
$t_{cav}$, between the flights of two successive pairs of atoms, the cavity
field interacts with its reservoir, and its equation of motion is given by~[6]
$$\dot{P}_n|_{loss} = 2(n+1)(\bar{n}_{th}+1)\kappa P_{n+1} -
2\kappa (n+ \bar{n}_{th} + 2n\bar{n}_{th})P_n + 2n\kappa \bar{n}_{th}P_{n-1}
\eqno(16)$$
where $\kappa = (2t_p)^{-1}$ is the cavity bandwidth, 
and $\bar{n}_{th}$ is the thermal
photon present in the cavity. Under the coarse-graining assumptions, the
complete equation of motion combines Eqs.(15) and (16) additively. This 
assumption has been seen to be valid~[20] for a random input of atoms
into the cavity, which is the case we are studying in this paper. We thus have,
$$\dot{P}_n = \dot{P}_n|_{gain} + \dot{P}_n|_{loss}\eqno(17)$$

\vskip 0.5in

\noindent
4. Steady-state photon statistics

\vskip 0.2in
 
The equation of motion for $P_n$, given by Eq.(17), involves $P_{n-2}$, and 
this makes it 
difficult to get a steady-state solution by the method followed in  case
of the one-atom micromaser~[6]. However, following Risken~[21],  Eq.(17) can
be recast in a tri-diagonal matrix equation involving the two-component
vector $T_n = \pmatrix{P_{2n} \cr P_{2n+1}}$, which provides an expression for
$T_n$ in the form of matrix continued fractions. Eq.(17) directly gives an
expression for $P_1$ in terms of $P_0$, and $P_0$ can be determined from
the normalization condition. This completes the determination of $P_n$.
But numerical evaluations of $P_n$ go out of control due to extremely slow
convergence of the matrix continued fractions.

Hence, we follow the method given below, which we find computationally
efficient. First, we set $n$ at a value $n = n_{max}$, say, and write
all the $P_n$ in a vector 
$$\vec{X} = \{P_0, P_1, P_2, .........., P_{n_{max}}\}^T\eqno(18)$$
This vector obviously has $n_{max}+1$ elements. We get $n_{max}$ equations
involving \\ 
$P_0,P_1,P_2,.......,P_{n_{max}}$ from Eq.(17). With the 
normalization condition
$$\sum_{n=0}^{n_{max}}P_n = 1\eqno(19)$$
we get the equation
$$M\vec{X} = \vec{I}\eqno(20)$$
where $M$ is a $(n_{max}+1)\times (n_{max}+1)$ matrix 
and $\vec{I}$
is another vector having again $n_{max}+1$ elements
$$\vec{I} = \{0,0,0,.........., 1\}^T\eqno(21)$$
The solution
$$\vec{X} = M^{-1}\vec{I}\eqno(22)$$
gives all $P_n$ for $0 \le n \le n_{max}$.

We analyze the photon statistics of the cavity field 
by numerically evaluating its
first and second moments, that is,
$$<n> = \sum_{n=0}^{n_{max}} nP_n\eqno(23)$$
proportional to the intensity of the cavity field and its normalized
variance
$$v = \sqrt{{<n^2> - <n>^2 \over <n>}}\eqno(24)$$
$v=1$ for a coherent state field, and thus, $v < 1$ indicates the nonclassical
nature of the cavity field. The process is repeated for a
higher value of $n_{max}$ and the corresponding $<n>$ and $v$ are compared
with those obtained from the earlier value of $n_{max}$. If they agree
to a desired accuracy, the process is stopped, and we have the
numerical values of the photon distribution function $P_n$, as well as the
average $<n>$ and variance $v$.

The results are displayed in Figs.1-3.
For the convenience of describing the 
micromaser action, we define the pump parameter
$$D = \sqrt{N}g\tau\eqno(25)$$
where $N = {\bar R}/2\kappa$ is the number of pairs of atoms 
that pass through the cavity
in a photon lifetime $(2\kappa)^{-1}$.  
The patterns in the variation of $<n>$
and $v$ as $D (\propto \tau)$ changes,  
can be understood from the dynamics~[22] of a collection
of two-level atoms interacting with the radiation field of a single mode
cavity with each atom being coupled to the cavity field by the
Jaynes-Cummings interaction~[2]. For a two-atom case with initial
condition $J=1$, it has been shown in [22] that the dynamics is controlled
by a d.c. term with a prefactor of the order of $P_n/n$ and terms 
oscillating at the first and second harmonic of the Rabi frequancy
$2g\sqrt{n+3/2}$ of the interaction Hamiltonian in Eq.(1c) and having
prefactors proportional to $P_n$ and $P_n/n$ respectively. In the
single-atom case~[23] (Jaynes-Cummings model), the well known dynamics
is controlled by only one term having the prefactor $P_n$ and oscillating
at the Rabi frequency $2g\sqrt{n+1}$ of the Jaynes-Cummings interaction~[2].
These differences make the present results different, in general, from
the one-atom micromaser action. 
Figs.1-3
show the differences clearly where we compare the results for the present
Dicke micromaser for $\bar{N}=100$ with the one-atom micromaser action~[4-6]
with a pump rate $\bar{N}=200$. The numerical values of the two pump rates
make the total number of atoms that pass through the two cavities having the
same $Q$ equal, and thus, justifies the comparison between the two systems.
Fig.2a shows that the superradiance nature of the interaction in Eq.2
makes the threshold values of $D$ lower compared to the one-atom micromaser.
The differences between the two systems become predominant for higher values
of $D$, that is, for longer interaction times. This is evidently due to the
cooperative interaction becoming more important for longer interaction times.
Thus, we notice in Fig.1  that $<n>$ is in general higher compared
to the one-atom micromaser for longer $\tau$.

The variance in the cavity field is in general different in the two systems.
It is generally very high near threshold, and hence, the sharp peaks in $v$
at threshold appear at different values of $D$ in the two systems as 
depicted in Fig.2b. We also notice in Fig.2b that the cavity field gets
nonclassical properties ($v<1$) for different ranges of $D$ in the two
systems. The photon distribution function also has different shapes as shown
in Figs.3, which clearly shows that the nature of the field depends clearly
on the nature of the pump. For example, in Fig.3a, the field in the one-atom
case is sub-Poissonian, while the field in the two-atom case is 
super-Poissonian. The field characteristics are vice-versa in Fig.3c.
However, the $P_n$ in the present case do not indicate any existence of the
so-called trapped states~[4-6].
The Rabi frequencies in the two-atom Dicke system, represented by the
eigen values $\lambda_0$, $\lambda_{+}$ and $\lambda_{-}$ 
do not provide clear conditions for the  trapped
states as one gets in the case of the one-atom micromaser~[6]. There the
conditions are essentially zeros of the function $Sin\sqrt{n+1}x$, $x$
being the dimensionless atom-field interaction time. The trapped states,
are however, washed out by occasional presence of two atoms in the cavity
of the one-atom maser~[7,8]. Thus the present analysis gives an easy
and transparent understanding of the disappearance of trapped states in
multi-atom events.
 
\vskip 0.2in

\noindent
5. Comparison with one-atom two-photon micromaser

The Dicke micromaser studied in the present paper has similarities with the
dynamics of a one-atom two-photon micromaser~[16] involving pumping of
atoms, individually into a microwave cavity where the atoms make a two-photon
transition~[19] from the upper to lower level via an intermediate level.
This is because the Dicke atomic system with $J=1$ is equivalent to a 
one-atom three-level system with the middle level having equal frequency
of separation $\omega$ from the upper and the lower levels. 
It may be noted here that $\omega$ is the transition frequency of the
individual two-level atoms in the Dicke system. If the cavity
eigenmode is in resonance with the two degenerate atomic transitions, and
in addition, if they have dipole moments of equal strength, then the present
theory can be employed to explain the dynamics of the two-photon cascade
micromaser. In other words, the results in Figs.1-3 can represent a one-atom 
two-photon cascade micromaser if we set the coupling constants of the two
degenerate transitions $g_1 = g\sqrt{2}$. The factor $\sqrt{2}$ in the Dicke
system represents its cooperative nature~[14].

We need not stop at this comparison. A Dicke micromaser with pumping of
three atoms at a time (all in their upper masing levels) will be equivalent
to a degenerate five-photon cascade micromaser, and so on.

However, a true two-photon process involves non-resonant one-photon
transitions. If $\omega_1$ and $\omega_2$ are two atomic transitions, and
$\nu$ is the cavity mode frequency, then a true two-photon process should
have large detunings $\triangle_1 = \omega_1 - \nu$ and $\triangle_2 = 
\nu - \omega_2$, that is, $\triangle_1$, $\triangle_2 \gg g_1$. 
A two-photon resonance
means $\triangle_1 + \triangle_2 = 0$, or $\omega_1 + \omega_2 = 2\nu$.
In the cascade two-photon micromaser discussed above, we have 
$\triangle_1 = \triangle_2 = 0$.
An extensive comparison of cavity-QED of cascade two-photon with true
two-photon processes can be found in [19] where
it has been shown that the two dynamics are, in general, different. 

The
present approach can also be applied to such a two-photon micromaser if
we can accomodate the one-photon detuning in the derivation of the photon
distribution function. This amounts to just setting the matrix elements
$[H_I]_{1,1} = [H_I]_{3,3}
= -\triangle$ in Eq.4. For simplicity, we set $g_1 = g\sqrt{2}$ here also. We
present the photon statistics for the two photon micromaser for various
values of $\triangle$ in Figs.4. We find that the operational characteristics
are strongly dependent on $\triangle$. We also notice that as $\triangle$
increases, the theory recovers the results for a two-photon micromaser
derived by using an effective two-level Hamiltonian, a derivation and
discussion of which can be found in [19].

For smaller values of $\triangle$, we notice wriggles in the variation 
of $<n>$ versus $D$ in Fig.4a. These wriggles begin at a value of
$D (\propto \tau)$ and persist as it is further increased. The value of
$D$ at which the wriggles begin to appear increases with $\triangle$. In other
words, the smooth variations in $<n>$
as in the case of an {\it effective two-level system}~[17,18],
occur for values of $D$ having an upper bound which increases with
$\triangle$. Such characteristics have been discussed in detail in the
context of the two-photon Jaynes-Cummings model in [19]. 
In addition to the above results, we also notice that the thresholds shift
to higher values of $D$ as $\triangle$ increases, and also, the range of
$D$ between two successive thresholds also increases with the one-photon
detuning. The cavity field usually has sub-Poissonian photon statistics
$(v < 1)$ between successive thresholds, and thus with a larger detuning,
this model provides sub-Poissonian fields for wider range of $D$ as shown
in Fig.4b.

\vskip 0.2in

\noindent
6. Conclusion

We have presented a theory for a two-atom Dicke micromaser and have brought
out the role played by the Dicke cooperativity on the micromaser action.
Though at first glance it may look that the Dicke micromaser results mimic
that for an equivalently pumped one-atom maser, the two systems are in
general different. At places, however, the two systems have similar trends,
an example of which can be seen in Fig.3b. We further show that the 
micromaser dynamics involving the three-level Dicke atomic system is formally
equivalent to a one-atom cascade two-photon micromaser. Interestingly, we note
that the two one-photon coupling constants in the two-photon cascade
transition has to scale $\sqrt{2}$ times of the 
one-atom coupling constants in the Dicke
system for the two micromaser actions to have identical results. As
mentioned above, this factor of $\sqrt{2}$ originates from the cooperative
nature of the Dicke system. 
 With the introduction of a one-photon detuning in the two-photon process, 
the present
approach also describes a one-atom two-photon micromaser. 
We have discussed in detail 
the role of the one-photon detuning in a two-photon micromaser action. 
Thus the Dicke superradiance effects can be quantitatively evaluated
by scaling, as mentioned above, the results from the one-atom two-photon
micromaser~[16].

Thus the Dicke micromaser model discussed in this paper is not beyond
the scope of experiments for verifying its results. The present paper
gives the clue to verify results from micromaser actions involving
pump mechanisms like the one studied here, which are rather difficult
to achieve experimentally. This suggests a way to experimentally
demonstrate the two-atom cavity-QED results~[22] in a one-atom
two-photon micromaser~[16]. It may be recalled that the one-atom
micromaser~[1] demonstrated the phenomenon of quantum revival~[23].
The same techniques may be employed in the case of the one-atom
two-photon micromaser~[16] and a proper scaling of its results
should demonstrate the two-atom cavity-QED results~[22].
 
\newpage
\noindent
REFERENCES

\begin{enumerate}

\item D. Meschede, H. Walther and G. Muller, Phys. Rev. Lett. {\bf 54},
551 (1985).
\item E. T. Jaynes and F. W. Cummings, Proc. IEEE {\bf 51}, 89 (1963).
\item G. Rempe, H. Walther and N. Klein, Phys. Rev. Lett. {\bf 58},
353 (1987).
\item P. Filipowicz, J. Javanainen and P. Meystre, Phys. Rev. A {\bf 34},
3077 (1986).
\item L. A. Lugiato, M. O. Scully and H. Walther, Phys. Rev. A {\bf 36},
740 (1987).
\item N. Nayak, Opt. Commun. {\bf 118}, 114 (1995).
\item M. Orszag, R. Ramirez, J. C. Retamal and C. Saavedra, Phys. Rev.
A {\bf 49}, 2933 (1994).
\item E. Wehner, R. Seno, N. Sterpi, B. -G. Englert and H. Walther,
Opt. Commun. {\bf 110}, 655 (1994).
\item G. M. D'Ariano, N. Sterpi and A. Zucchetti, Phys. Rev. Lett. {\bf 74},
900 (1995).
\item M. Elk, Phys. Rev. A {\bf 54}, 4351 (1996).
\item M. I. Kolobov and F. Haake, Phys. Rev. A {\bf 55}, 3033 (1997).
\item K. An, J. J. Childs, R. R. Dasari and M. S. Feld, Phys. Rev. Lett.
{\bf 73}, 3375 (1994).
\item H. J. Carmichael and B. C. Sanders, Phys. Rev. A {\bf 60}, 2497
(1999). 
\item R. H. Dicke, Phys.  Rev. {\bf 93}, 99 (1954).
\item J. M. Raimond, P. Goy, M. Gross, C. Fabre and S. Haroche, Phys. Rev.
Lett. {\bf 49}, 1924 (1982).
\item M. Brune, J. M. Raimond, P. Goy, L. Davidovich and S. Haroche,
Phys. Rev. Lett. {\bf 59}, 1899 (1987).
\item L. Davidovich, J. M. Raimond, M. Brune and S. Haroche, Phys. Rev.
A {\bf 36}, 3771 (1987).
\item I. Ashraf, J. Gea-Banacloche and M. S. Zubairy, Phys. Rev. A {\bf 42},
6704 (1990).
\item V. Bartzis and N. Nayak, J. Opt. Soc. Am. B {\bf 8}, 1779 (1991).
\item J. Bergou, L. Davidovich, M. Orszag, C. Benkert, M. Hillary and
M. O. Scully, Phys. Rev. A {\bf 40}, 5073 (1989); L. Davidovich,
S. -Y. Zhu, A. Z. Khoury and C. Su, Phys. Rev. A {\bf 46} 1630 (1992).
\item H. Risken, {\it Fokker - Planck Equations} (Springer - Verlag,
Berlin, 1984), p. 200.
\item G. Ramon, C. Brif and A. Mann, Phys. Rev. A {\bf 58}, 2506 (1998).
\item N. Nayak, R. K. Bullough, B. V. Thompson and G. S. Agarwal, IEEE
J. Quant. Electron QE {\bf 24}, 1331 (1988) and references therein.
\end{enumerate}

\newpage

\noindent
FIGURE-CAPTIONS

\vskip 0.2in

\noindent
Fig.1 

The cavity field intensity, proportional to $<n>$, versus the pump parameter
$D$. $\bar{n}_{th} = 0.1$. $\bar{N}=100$ and $200$ for the Dicke micromaser
(a) and one-atom micromaser (b) respectively. This makes the total number
of atoms that pass through the cavity in the two systems equal. The curve
(b) is shifted upwards by $150$ for clarity.

\vskip 0.2in

\noindent
Fig.2a

Variation in $<n>$ with respect to $D$ for shorter interaction time. The
parameters are same as in Fig.1. The full and broken curves are for the
Dicke micromaser and one-atom micromaser respectively.

\vskip 0.2in

\noindent
Fig.2b

Variation of $v$ with $D$. The other parameters are same as in Fig.2a.

\vskip 0.2in

\noindent
Fig.3a

Photon distribution function for $D=25$. The other parameters are same as
in Figs.2. $P(n)$ is sub-Poissonion $(v=0.60761)$ in the case of one-atom
micromaser (broken curve), and is super-Poissonian $(v=1.27596)$ in the
case of Dicke micromaser.

\vskip 0.2in

\noindent
Fig.3b

$P(n)$ for $D=50$. The other parameters are same as in Fig.3a. For both the
cases, $P(n)$ is super-Poissonian at this value of $D$. $v=1.09944$ (broken
curve) and $v=1.15871$ (full curve).

\noindent
Fig.3c

$D=400$. $P(n)$ is sub-Poissonian $(v=0.22911)$ in the Dicke micromaser
system (full curve). One-atom micromaser (broken curve) provides a 
super-Poissonian $(v=1.05726)$ in this case.

\vskip 0.2in

\noindent
Fig.4a

$<n>$ versus $D$ for a two-photon micromaser for three different values
of the one-photon detuning $\triangle$. $\bar{n}_{th}=0.1$ and $\bar{N}=100$.
$\triangle = 100$ (curve (a)), $\triangle = 150$ (curve (b)), and
$\triangle = 300$ (curve (c)). For clarity, the curves (b) and (c) are
shifted upwards by $40$ and $80$ respectively.

\vskip 0.2in

\noindent
Fig.4b

Variation of $v$ with $D$ for a two-photon micromaser. The other parameters
are same as in Fig.4a. Here, the curves (b) and (c) are shifted by $4$ and
$8$ respectively, for clarity.

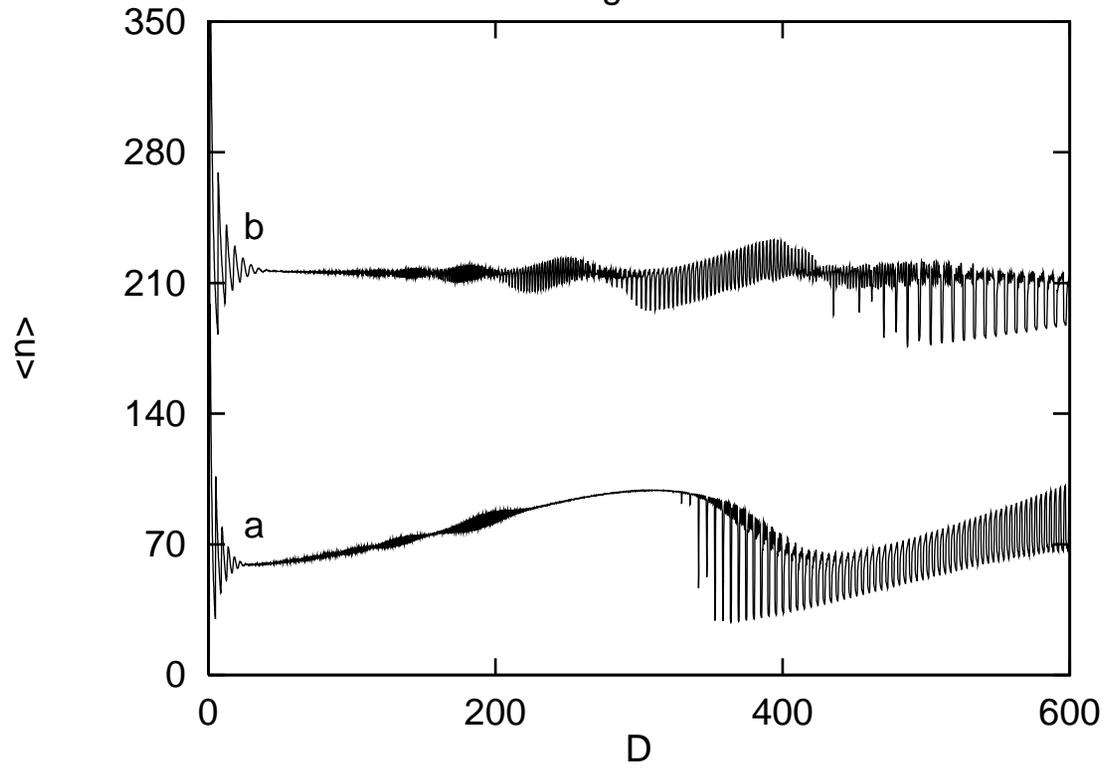
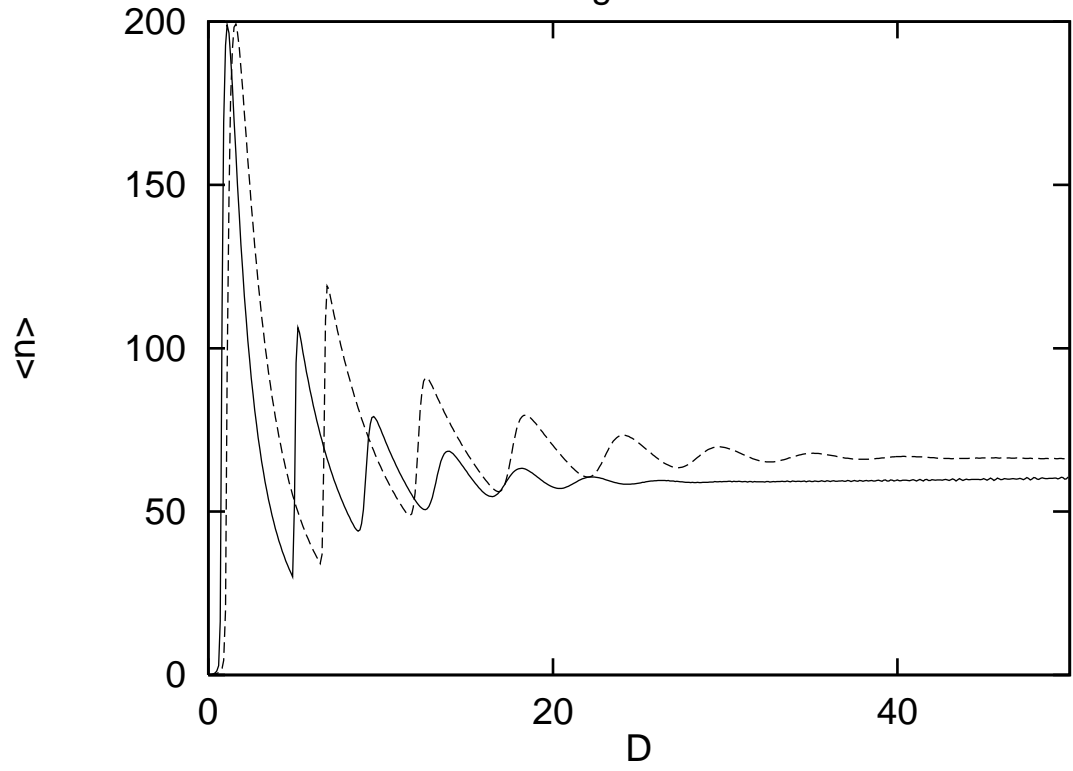
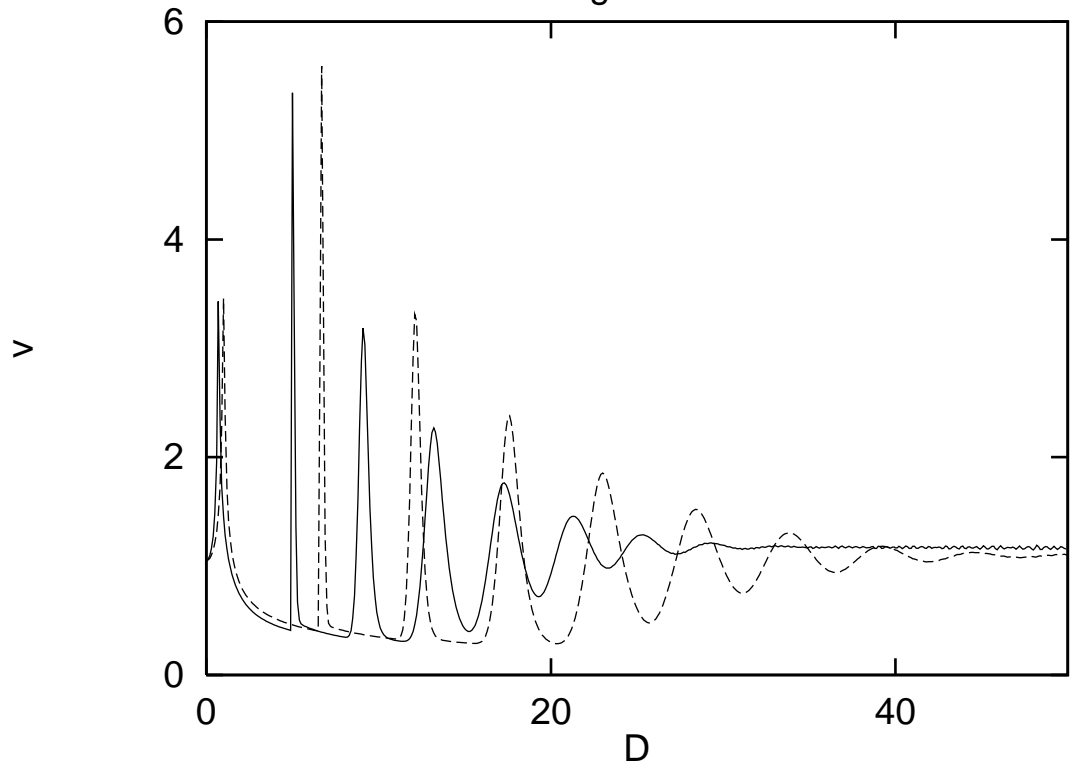
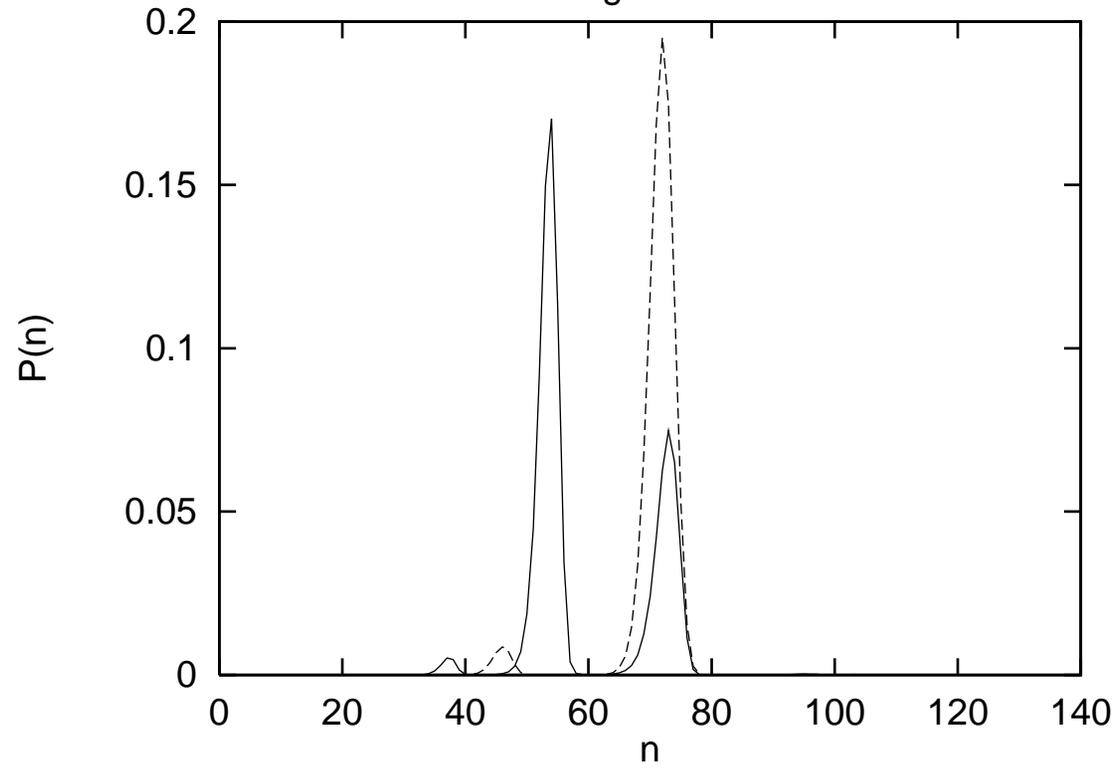
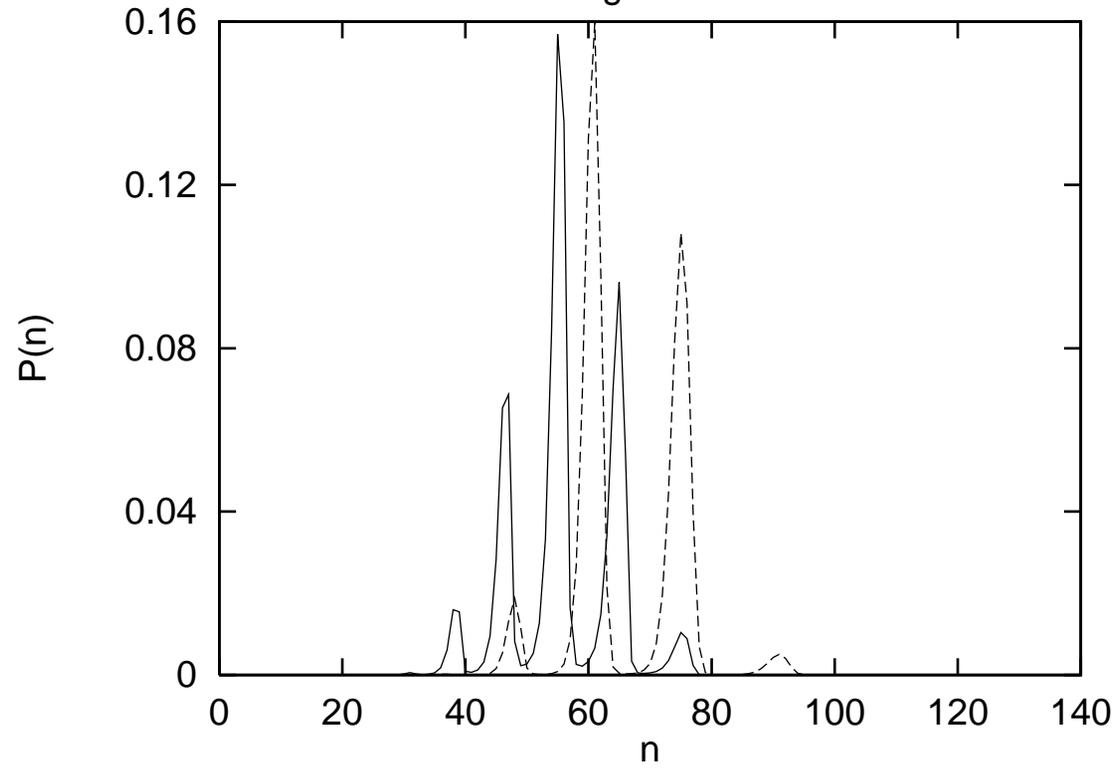
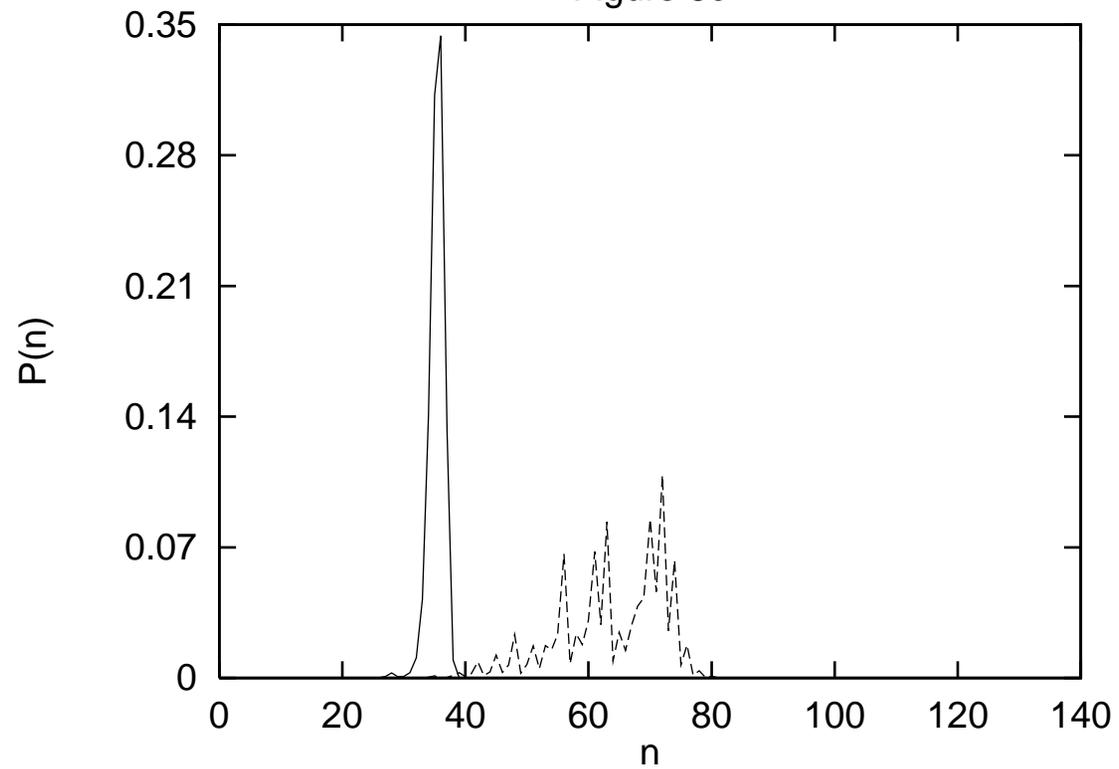
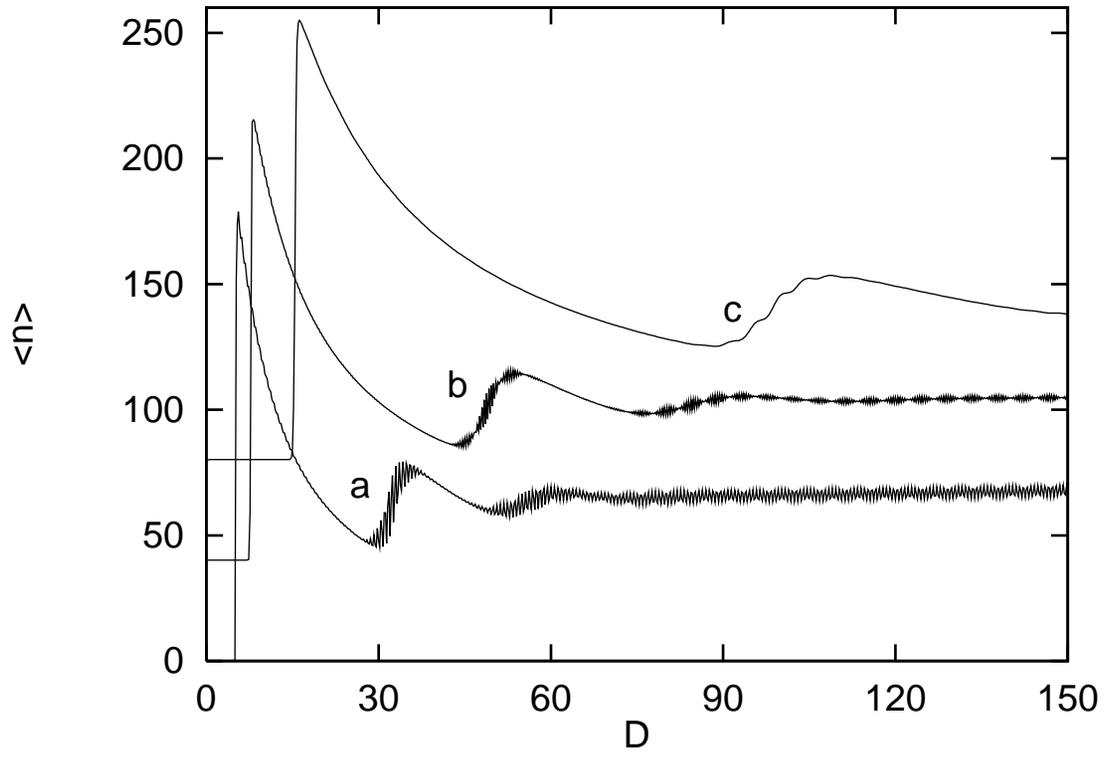
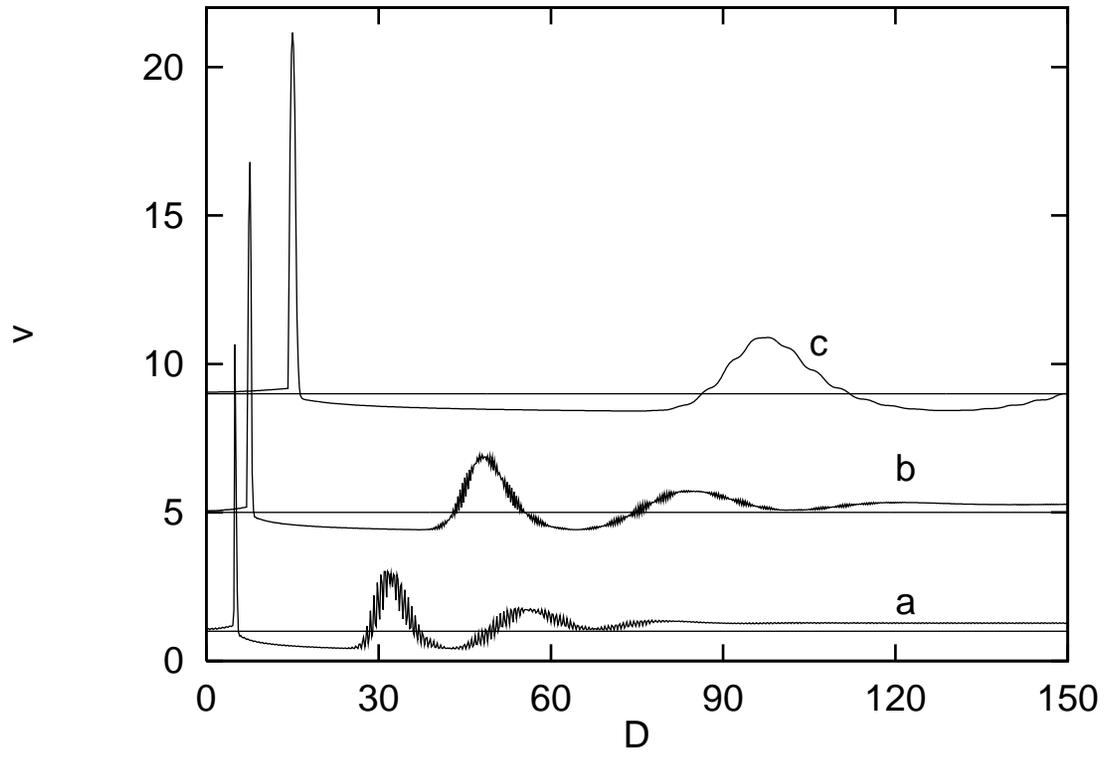
\end{document}